\documentclass[10pt,twocolumn]{article}
\usepackage[top=1.9cm, bottom=1.9cm, left=1.9cm, right=1.9cm]{geometry}
\usepackage{times}  %
\usepackage[hyphens]{url}
\usepackage{graphicx}  %
\usepackage[keeplastbox]{flushend}
\usepackage[hyphens]{url}  %
\usepackage{graphicx} %
\usepackage{tikzsymbols}

\usepackage{url}                                %
\usepackage{array,multirow,graphicx,adjustbox}  %
\usepackage{booktabs}                           %
\usepackage[utf8]{inputenc}
\usepackage[ruled,algosection,noend,linesnumbered]{algorithm2e}
\usepackage{float}
\usepackage{paralist}
\usepackage{amsmath}
\usepackage{amsfonts}
\usepackage{xcolor}
\usepackage{csquotes} 
\usepackage{tabularx}
\usepackage{makecell}
\usepackage{pbox}
\usepackage[hang,flushmargin]{footmisc}
\usepackage{flushend}
\usepackage{xspace}
\usepackage{multicol, lipsum}
\usepackage[T1]{fontenc}

\usepackage{xcolor}
\usepackage{booktabs}
\usepackage{graphicx}
\usepackage{paralist}
\usepackage[small,bf]{caption}
\usepackage[tight]{subfigure}
\usepackage[hang,flushmargin]{footmisc}

\usepackage[compact]{titlesec}
\titlespacing*{\section}{0pt}{*4}{4pt}
\titlespacing*{\subsection}{0pt}{*3}{3pt}
\usepackage{xspace}

\makeatletter
\def\url@leostyle{%
  \@ifundefined{selectfont}{\def\UrlFont{}}%
  {\def\UrlFont{}}%
}
\makeatother
\usepackage[hyphenbreaks]{breakurl}

\usepackage[bookmarks=true, bookmarksnumbered=true, colorlinks=true, linkcolor=linkcol, citecolor=citecol, urlcolor=urlcol, hypertexnames=true]{hyperref}

\definecolor{darkgreen}{RGB}{0, 100, 0}
\definecolor{linkcol}{rgb}{0.3,0,0}
\definecolor{citecol}{rgb}{0.3,0,0}
\definecolor{urlcol}{rgb}{0.3,0,0}

\makeatletter
\def\url@leostyle{%
  \@ifundefined{selectfont}{\def\UrlFont{\small}}%
  {\def\UrlFont{}}%
}
\makeatother
\newcommand{\descr}[1]{\smallskip\noindent\textbf{#1}}
\newcommand{\politics}{{{\selectfont /v/politics}}\xspace}
\newcommand{\dev}{{{\selectfont /v/voatdev}}\xspace}

\newcommand{\news}{{{\selectfont /v/news}}\xspace}
\newcommand{\world}{{{\selectfont /v/WorldToday}}\xspace}
\newcommand{\travel}{{{\selectfont /v/travel}}\xspace}
\newcommand{\television}{{{\selectfont /v/television}}\xspace}
\newcommand{\fph}{{{\selectfont /v/fatpeoplehate}}\xspace}
\newcommand{\coon}{{{\selectfont /v/CoonTown}}\xspace}
\newcommand{\announcements}{{{\selectfont /v/announcements}}\xspace}
\newcommand{\nigger}{{{\selectfont /v/Nigger}}\xspace}
\newcommand{\redpill}{{{\selectfont /v/TheRedPill}}\xspace}
\newcommand{\deep}{{{\selectfont /v/DeepFake}}\xspace}

\newcommand{\videos}{{{\selectfont /v/videos}}\xspace}
\newcommand{\funny}{{{\selectfont /v/funny}}\xspace}
\newcommand{\whatever}{{{\selectfont /v/whatever}}\xspace}
\newcommand{\askvoat}{{{\selectfont /v/AskVoat}}\xspace}
\newcommand{\greatawakening}{{{\selectfont /v/GreatAwakening}}\xspace}
\newcommand{\thegreatawakening}{{{\selectfont /v/TheGreatAwakening}}\xspace}

\newcommand{\pizzagate}{{{\selectfont /v/pizzagate}}\xspace}

\newcommand{\theawakening}{{{\selectfont /v/theawakening}}\xspace}

\newcommand{\neofag}{{{\selectfont /v/NeoFAG}}\xspace}

\newcommand{\qrv}{{{\selectfont /v/QRV}}\xspace}

\newcommand{\meanwhileonreddit}{{{\selectfont /v/MeanwhileOnReddit}}\xspace}

\newcommand{\rfph}{{{\selectfont /r/fatpeoplehate}}\xspace}
\newcommand{\rnigger}{{{\selectfont /r/nigger}}\xspace}
\newcommand{\rdeep}{{{\selectfont /r/DeepFakes}}\xspace}
\newcommand{\rpizzagate}{{{\selectfont /r/pizzagate}}\xspace}
\newcommand{\rincel}{{{\selectfont /r/incel}}\xspace}
\newcommand{\rbeatingwomen}{{{\selectfont /r/beatingwomen}}\xspace}

\newcommand{\rcbts}{{{\selectfont /r/CBTS\_Stream}}\xspace}
\newcommand{\rthefappening}{{{\selectfont /r/TheFappening}}\xspace}
\newcommand{\rgreatawakening}{{{\selectfont /r/GreatAwakening}}\xspace}

\let\OLDthebibliography\thebibliography
\renewcommand\thebibliography[1]{
  \OLDthebibliography{#1}
  \setlength{\parskip}{0pt}
  \setlength{\itemsep}{1pt plus 0.2ex}
}

\newif
\ifcomment
\commenttrue
\ifcomment
\newcommand{\sz}[1]{{\bf \textcolor{brown}{SZ: #1}}}
\newcommand{\edc}[1]{{\bf \textcolor{red}{EDC: #1}}}
\newcommand{\gs}[1]{{\bf \textcolor{green}{GS: #1}}}
\newcommand{\ap}[1]{{\bf \textcolor{blue}{AP: #1}}}
\newcommand{\jbnote}[1]{{\bf \textcolor{magenta}{JB: #1}}}
\else
\newcommand{\sz}[1]{}
\newcommand{\edc}[1]{}
\newcommand{\gs}[1]{}
\newcommand{\ap}[1]{}
\newcommand{\jbnote}[1]{}
\fi

\title{``I Can’t Keep It Up.''\\A Dataset from the Defunct Voat.co News Aggregator}
\author {
    Amin Mekacher, Antonis Papasavva \\
    City, University of London, 
    University College London \\
    amin.mekacher@city.ac.uk, antonis.papasavva@ucl.ac.uk\\
    \normalsize --- iDRAMA Lab, https://idrama.science ---
    \\
    \\*Please cite the ICWSM 2022 version of the paper*
}
\date{}

\begin{document}

\maketitle

\begin{abstract}
Voat.co was a news aggregator website that shut down on December 25, 2020. 
The site had a troubled history and was known for hosting various banned subreddits. 
This paper presents a dataset with over $2.3M$ submissions and $16.2M$ comments posted from $113K$ users in $7.1K$ \emph{subverses} (the equivalent of subreddit for Voat). 
Our dataset covers the whole lifetime of Voat, from its developing period starting on November 8, 2013, the day it was founded, April 2014, up until the day it shut down (December 25, 2020).

This work presents the largest and most complete publicly available Voat dataset, to the best of our knowledge. 
Along with the release of this dataset, we present a preliminary analysis covering posting activity and daily user and subverse registration on the platform so that researchers interested in our dataset can know what to expect.

Our data may prove helpful to false news dissemination studies as we analyze the links users share on the platform, finding that many communities rely on alternative news press, like Breitbart and GatewayPundit, for their daily discussions. 
In addition, we perform network analysis on user interactions finding that many users prefer not to interact with subverses outside their narrative interests, which could be helpful to researchers focusing on polarization and echo chambers.
Also, since Voat was one of the platforms banned Reddit communities migrated to, we are confident our dataset will motivate and assist researchers studying deplatforming. 
Finally, many hateful and conspiratorial communities were very popular on Voat, which makes our work valuable for researchers focusing on toxicity, conspiracy theories, cross-platform studies of social networks, and natural language processing.

\end{abstract}

\section{Introduction}
Social networks are a primary tool in today's society.
They offer countless opportunities for people around the world to connect in various ways, find jobs, entertain themselves, catch up on world happenings, etc.
At the same time, social networks sometimes offer a ``safe-house'' for people that want, among other things, to connect to like-minded individuals towards sharing hate and toxicity~\cite{almerekhi2020investigating,vice2017toxicreddit}, discussing controversial matters~\cite{time2021racismtwitter}, and spreading misinformation and disinformation~\cite{mit2021fbmisinformation}.

Mainstream social networks suffer from users and communities that organize these conversations on their platforms.
A common ``solution'' the administrators result to is to ban these users - \emph{deplatforming}.
The social network that is known to have taken this action many times so far is Reddit, which banned more than $7K$ subreddits~\cite{verge2020redditnans} from its platform - the first one being in 2014~\cite{daily2014firstban}. 
Research on deplatforming shows that users that had their communities banned met on other platforms and forums and even got more toxic than what they used to be~\cite{horta2021platform}.
Other than forums, users move to social networks that allow controversial discussions.
One of the platforms that many banned Reddit communities decided to migrate to was Voat.

Voat was a Reddit-esque social network founded in April 2014 and shut down in December 2020~\cite{verge2020whatisvoat}.
Similar to Reddit, discussions on Voat are divided into various channels - \emph{subverses} - the equivalent of a subreddit. 
Users can subscribe to as many subverses they wish but cannot moderate more than ten to prevent users gaining undue influence on the platform.
Registration of new users on Voat requires only a unique username and a password.
Newcomers can upvote, downvote, and comment on existing submissions but cannot create new submissions under subverses until they achieve a certain amount of upvotes on all of their comments.

Since its foundation, Voat gradually gained popularity over the years, especially after every Reddit cleansing~\cite{verge2015migration,times2016migration,daily2017migration,fox2018migration}.
Overall, Voat is known for hosting banned extreme communities and users, providing a safe space for like-minded individuals to share their ideas ``freely.''
Voat has attracted the interest of researchers before as it hosted communities like \fph, \coon, and \nigger~\cite{chandrasekharan2017bag}, \redpill~\cite{saleem2017web}, \greatawakening~\cite{papasavva2021qoincidence}, etc.

\descr{Data Release.}
In this work, we present, to the best of our knowledge, the largest and most complete dataset of Voat.
Along with this paper, we release a dataset~\cite{zenodo} that consists of over $18.6M$ posts from $113K$ users in $7K$ subverses over the lifetime of Voat (November 2013 - December 2020).
Specifically, our dataset is four fold:
\begin{compactitem}[--]
\item{Title, body, and metadata of submissions;}
\item{content and metadata of comments;}
\item{user profile data; and}
\item{subverse profile data.}
\end{compactitem}

\descr{Relevance.} 
Our dataset provides several opportunities to the research community.
First, Voat was evidently the place many banned users and communities moved to after being banned on other platforms~\cite{papasavva2020raiders,chandrasekharan2017bag}.
To this end, our dataset can assist researchers that focus on deplatforming and user migration. 
Also, our dataset may aid researchers deepen our understanding of how and when these communities choose their new ``home'' after being removed from their previous one.
Second, our dataset covers numerous offline events like the 2016 and 2020 US Presidential Election and debate, Brexit, Epstein's arrest, and various terrorist attacks and unrest around the world that can prove helpful in further analysis of these events. 
Third, since Voat was a supporter of freedom of expression and online free speech for extreme and hateful communities, it contains a variety of slang language and toxic content that can be useful towards understanding hateful communities.

\descr{Paper organization.}
The rest of the paper is organized as follows. 
First, we briefly explain what Voat is and how it works in Section~\ref{sec:whatisvoat} before going through its history in Section~\ref{sec:history}.
Then, we describe the process of parsing the Internet Archive Wayback Machine (IAWM) published data, along with the complimentary collection of additional user and subverse data in Section~\ref{sec:collection}.
We then describe the structure of our dataset in Section~\ref{sec:data_description} and provide a statistical analysis of the dataset (Section~\ref{sec:analysis}), followed by reviewing related work (Section~\ref{sec:related_work}).
The paper concludes with Section~\ref{sec:conclusion}.

\section{What \emph{was} Voat?}\label{sec:whatisvoat}
Voat was a Reddit-esque news aggregator launched in April 2014.
The mascot of Voat resembles an angry goat, which was designed and freely offered to the website by a user of the site.

\descr{Subverses.}
Discussions on Voat occur in specific groups of interests called ``subverses.''
Users could create subverses on-demand before June 2020, when the administrators disabled this functionality.
When a user creates a subverse, they become its owner, hence having complete authority over the subverse: they can deactivate the subverse and appoint other co-owners and moderators.
The moderators can delete submissions and comments posted by users and even ban users from posting on the subverse.
The owners and moderators can also allow users to post anonymously in their subverse, which replaces the posters' username with a random multi-digit number; not unique to each user.
To prevent users from gaining extreme influence on the platform, Voat limits the number of subverses one can own or moderate.%

\descr{Users.}
Voat proclaimed itself as a free-speech platform that offered its users anonymity.
When newcomers register a new account, Voat does not require any personal details to verify the account, like an email address or phone number.
A user can insert a username and a password to register, but if they forget their password, there is no way to recover the account.
 
After registering a new account, users can subscribe to subverses of interest, comment, upvote, and downvote the comments and submissions but cannot post new submissions.
To post a new submission, they first need to acquire ten Comment Contribution Points (CCP).
To do so, newcomers post comments on existing submissions trying to collect a \emph{net score} of ten upvotes on all of their comments (one downvote cancels one upvote).
The privilege of posting submissions is not guaranteed as users may lose it if their CCP falls below ten.
Although this functionality may discourage users from being toxic to each other, it might also prevent users from debating their opinions as others may disagree and downvote them.
Voat users often refer to themselves as ``goats'' due to the platform's mascot.

\descr{Submissions and voting system.}
Voat was a news aggregator platform, hence users can create a new submission by posting a title and a description, accompanied by a link to a news source, optionally. %
If the poster provides a link, the submission's title becomes a hyperlink to the source website. 
The domain of the source website appears next to the submission's title, along with the poster's username, and the date and time the submission was posted.

Similar to Reddit, Voat offers a hierarchical, tree-like commenting system: other users can comment on the submission and the comments of other users. 
Users can upvote or downvote the submission or other users' comments.
In contrast with Reddit, Voat displays the total number of upvotes and downvotes a submission or a comment received.
Also, the downvote functionality on Voat is not the same as Reddit's: downvoted submissions and comments alert the moderators of spammy or illegal content so they can take action.
This functionality enforces the establishment of echo chambers as users usually downvote content that does not align with their beliefs.
This usually results in the downvoted user to either losing their submission posting privilege or even being banned from the subverse.

\descr{Content visibility.}
Voat attempted to provide its users with some ephemerality without deleting its content, but hiding it instead.
Voat subverses filter submissions under three tabs, namely, hot, new, and top.
Each subverse has 500 active submissions in 20 pages (0 to 19).
Hot submissions are the ones that are currently active and discussed, new submissions are the ones that were posted most recently, and top submissions are the most popular submissions of the subverse (many comments).
Many subverses disabled the functionality of these tabs, and the submissions shown across all three tabs are often the same, just in a different order.
We note that our dataset does not contain the tab of submission's as tabs are merely filters and often change based on the status of the submission, e.g., from new to hot.

When a user creates a new submission on a subverse, it would typically appear first on the new tab on page 0. 
At the same time, the last submission of page 19 is archived but not deleted, meaning if one knows the link to that submission they can still reach it but cannot comment or vote it.

\descr{Voat API.}
Voat supported a JSON API service for some time, but its maintenance stopped in October 2020.
To collect the submissions of a subverse, one had to request the API of a specific page number (0 to 19) of a subverse's tab.
The response of the API would be the 25 submissions of that page without their comments.
To collect the comments, one needs to request them using the submission ID number, in which the API responds with 25 comments at a time.

Thus, to collect all the submissions from a subverse, one needs to request all 20 pages for the three tabs separately from the API.
As explained by~\cite{papasavva2021qoincidence}, the API does not list the archived subverses and does not respond to requests where the page is above 19.
However, if one knows the submission ID and the subverse it was posted in, they can request the API for that specific submission.
Since submission IDs on Voat are incremental, one could theoretically collect all of Voat's submissions by requesting the API for each submission ID incrementally for more than 7.5K subverses; that is 7.5K requests, in the worst case, to collect a single submission.
To the best of our knowledge, no study or work managed to collect the full Voat dataset.

\descr{SearchVoat.}
A website not associated with Voat, named searchvoat.co, used to collect the Voat submissions and comments for its users to browse.\footnote{\url{https://searchvoat.co/search.php}}
The site does not support an API and does not allow web scraping.
After Voat shut down, the site became a news aggregator, similar to Voat.\footnote{\url{https://searchvoat.co/forum/}}

\section{Voat's Troubled History}\label{sec:history}
In this section, we present Voat's history as we believe it highlights the significance of our dataset.
WhoaVerse was the original name of the website and it was founded in April 2014.
The website was a hobby project of Atif Colo (Voat username @Atko). 
Justin Chastain later joined Colo as a co-founder (username @PuttItOut).
The founders advertised the website as an alternative social network focusing on freedom of expression and speech, which satisfies its users' needs and wants. 
In December 2014, WhoaVerse changed its name to Voat and marked its mascot as an angry goat.

In June 2015, after Reddit banned various hateful subreddits~\cite{verge2015migration}, including \rnigger and \rfph, many Reddit users started registering accounts on Voat. 
The sudden influx of users overloaded the site, causing temporary down time~\cite{voat2015influx}. 

On June 19, 2015, Voat's web hosting service, Host Europe,\footnote{\url{https://www.hosteurope.de/en/}} canceled Voat's contract claiming that the site is publicizing abusive, insulting, youth-endangering content, along with illegal right-wing extremist content~\cite{host2015ban}.
Some days later, PayPal froze Voat's payment processing services~\cite{paypal2015ban}. 
In response, Voat shut down four subverses, two of which hosted sexualized images of minors and the founders attributed the shutdown to political correctness~\cite{jailbait}. 
The site moved to a different hosting provider and started accepting cryptocurrency donations. 

In July 2015, Reddit banned a popular administrator that caused another influx of Reddit members registering with Voat, leading to more downtime.
In an interview, Colo said that they ``provide an alternative platform where users would not be censored and still say whatever they want''~\cite{interview2015colo}.
Voat was the target of DDoS attacks many times and experienced numerous failures during its six years of operation.
The most significant attack was in July 2015~\cite{ddos2015}. 
Voat, Inc. became a registered corporation in the US in August 2015.
Although Voat was based in Switzerland, the U.S. seemed like the best option as explained by Colo in a post: ``US law with regards to free speech, by far beats every other candidate country we've researched.''

In November 2016, more users relocated to Voat after Reddit banned the \rpizzagate conspiracy theory subreddit~\cite{times2016migration}. 
In January 2017, Colo resigned as CEO of Voat due to time availability restrictions and was replaced by Chastain.
Chastain ran a fundraiser campaign in May 2017 after announcing that Voat might have to shut down due to financial issues; Voat managed to stay online.

In November 2017, Reddit banned its incel community (\rincel), and many of its followers reportedly moved to Voat~\cite{daily2017migration}. 
About a year later, on September 12, 2018, Reddit banned numerous subreddits dedicated to the QAnon conspiracy theory, which again caused many QAnon adherents to migrate to Voat~\cite{papasavva2021qoincidence}.

In April 2019, Voat's CEO Chastain asked Voat users to stop threatening people as he had been contacted by a ``US agency'' about the threats posted on the website.\footnote{\url{https://searchvoat.co/v/Voat/3178819}} 
In response, Voat users were not pleased to hear that Voat was working with agencies to remove Voat content and ``limiting'' the site's freedom of expression.
Specifically, the first comment on the submission was an anti-Semitic slur, calling for the extermination of Jews~\cite{agency2019}.

Finally, on December 22, 2020, Voat announced again, now for the last time, that it would shut down due to lack of funding.\footnote{\url{https://searchvoat.co/v/announcements/4169936}}
Chastain explained that he had been funding the site himself since March 2020 but had run out of money.
On December 25, 2020, Voat shut down and its last submission was posted by Chastain, noting: ``@Atko made the first post to Voat, so I am making the last.''\footnote{\url{https://searchvoat.co/v/Voat/4174956}}

In Table~\ref{tbl:events}, we list some aforementioned Reddit bans that probably affected Voat's activity.
Some of these bans previously captured researchers' interest.
We use these bans in our analysis in Section~\ref{sec:analysis} to show whether Voat's activity was indeed affected. 

\begin{table}[htbp]
\centering
\resizebox{1\columnwidth}{!}{
\begin{tabular}{llr}
\toprule
\textbf{No} & \textbf{Date} & \textbf{Ban}   \\
\hline
1 & May 9, 2014 & \rbeatingwomen \cite{fappeningbans} \\
2 & Sep 6, 2014 & \rthefappening \cite{fappeningbans} \\
3 & May 7, 2015 & \rnigger \cite{chandrasekharan2017bag} \\
4 & Jun 6, 2015 & \rfph \cite{chandrasekharan2017bag} \\
5 & Nov 23, 2016 & \rpizzagate \cite{times2016migration}\\
6 & Nov 7, 2017 & \rincel \cite{daily2017migration} \\
7 & Mar 15, 2018 & \rcbts \cite{fox2018migration} \\
8 & Sep 18, 2018 & \rgreatawakening \cite{papasavva2021qoincidence} \\
\bottomrule
\end{tabular}}
\caption{Reddit bans that reportedly affected Voat's activity.}
\label{tbl:events}
\end{table}

\begin{table}[htbp]
\small
\centering
\begin{tabular}{lrrr}
\toprule
\multicolumn{1}{c}{\textbf{}} & \textbf{Count}       & \textbf{\# Users}           & \textbf{\# Subverses}     \\ 
\toprule
\textbf{Submissions}          & 2,334,817            & \multicolumn{1}{r}{80,063}  & \multicolumn{1}{c}{7,616} \\
\textbf{Comments}             & 15,731,754          & \multicolumn{1}{c}{153,827} & \multicolumn{1}{c}{7,515} \\
\textbf{Subverses}            & 7,094                & \multicolumn{1}{c}{}        & \multicolumn{1}{c}{}      \\
\textbf{Users}                & 108,451              & \multicolumn{1}{c}{}        & \multicolumn{1}{c}{}      \\
\toprule
\end{tabular}
\caption{Number of submissions, comments, user profiles, and subverse profiles in the IAWM dataset.}
\label{tbl:data_crawled}
\end{table}

\begin{table}[htbp]
\centering
\small
\resizebox{\columnwidth}{!}{\begin{tabular}{lrrrr}
\toprule
\textbf{} & \textbf{Submissions} & \textbf{Comments}  & \textbf{Users}    & \textbf{Subverses} \\ 
\toprule
\textbf{Total} & 2,380,262 & 16,263,309 & 113,431 & 7,095 \\
\toprule
\end{tabular}}
\caption{Final number of submissions, comments, user profiles, and subverse profiles in the released dataset.}
\label{tbl:final_dataset}
\end{table}

\section{Data Parsing and Data Collection}\label{sec:collection}
This section details the methodology and tools employed for our data collection infrastructure. 

\descr{Submissions and Comments}.
Following Voat's shutdown on December 25, 2020, the Archive Team\footnote{For more details about the Archive team, see \url{wiki.archiveteam.org}} released a set of Voat snapshot captures in Web ARChive (WARC) format~\cite{archive}, hosted on the Internet Archive Wayback Machine (IAWM).
These WARC captures include snapshots the IAWM captured over the lifetime of Voat. 
A WARC format file consists of single or multiple WARC records (snapshots), and it supports, among other things, the access and scraping of archived data.
The files also hold revised and duplicated snapshots~\cite{warc}.

To parse these snapshots into structured data, we set up a Python script to parse the submissions and comments.
In our case, every WARC file is a collection of various Voat snapshots the IAWM captured.
To facilitate the smooth parsing of the WARC files, we use the \emph{warcio} Python library.\footnote{\url{https://pypi.org/project/warcio/}}
This library offers a convenient and reliable way to read a WARC file by streaming every entry included in the file and automatically detecting the \emph{payload}.
The payload contains the capture itself, i.e., the HTML DOM tree code of the platform.
Each WARC file includes the snapshot of the entire platform for a specific time and date, that is, thousands of submission pages for millions of submissions. 

Our parser captures the HTML DOM tree code of each page included in the WARC files serially.
Then, it passes the HTML DOM tree to a function that uses the \emph{beautifulsoup} Python library to read and store in JSON format the data and metadata of the submissions and comments, i.e., submission title and content, number of upvotes and downvotes, comments, etc.\footnote{\url{https://pypi.org/project/beautifulsoup4/}}
We ensure that our parser only stores the latest submission version, as WARC files have duplicate data. 

We note that although many languages appear in our dataset, the overwhelming majority of posts use the English language.
In addition, our parser does not capture or store any visual media, like videos and pictures, since such files are not included in the snapshots.
Hence, our dataset is not suitable for researchers focusing on visual media analysis.

\descr{User and subverse profiles.}
To complement our dataset, we also collected user and subverse profiles.
A user profile includes data like username and registration date, whereas a subverse profile consists of data like subverse creation date, description, etc.
To collect this data, we built a crawler using the IAWM API,\footnote{\url{https://pypi.org/project/waybackpy/}} beautifulsoup, and HTML requests.\footnote{\url{https://pypi.org/project/requests-html/}}

Every user and subverse profile URL is unique, but they all start the same way: \emph{voat.co/u} for the former and \emph{voat.co/v} for the latter.
First, we request the IAWM API for all the snapshots whose URLs start like user or subverse URLs.
We then collect the responses and parse them into JSON format, storing the latest snapshot the IAWM has in its database for every unique username and subverse profile URL.

The above process results in the dataset summarized in Table~\ref{tbl:data_crawled}.
We collect a dataset that consists of more than $2.3M$ submissions posted by $80K$ users in $7.6K$ subverses, and over $15.7M$ comments posted by almost $154K$ users.
Note that IAWM does not have the profiles of about 500 subverses and hence we only manage to collect the profiles of $7.1K$ subverses ($6.8\%$ loss).
In addition, we collect almost $108.5K$ unique user profiles.

\descr{Data collected via Voat API.}
In an attempt to complete our dataset, we merge it with the data collected for the~\cite{papasavva2021qoincidence} study.
For that study, we collected $176K$ submissions and $1.45M$ comments posted from $28K$ users in $241$  subverses.
Our data collection infrastructure used Voat's API between May 2020 and October 2020, when Voat stopped the maintenance of its API.
We find $45.5K$ submissions and $532K$ comments that were missing from the IAWM archive and incorporate them in the released dataset.

Some subverses on Voat offered anonymity to their users by replacing their username with a random eight-digit number (not a unique number for every user).
The total number of users that commented or posted a submission (Table~\ref{tbl:data_crawled}) does not include anonymous or deleted users.
Hence, we assume that Voat's known user base is $155K$ users, at least, based on the data we collect from the IAWM.
It is impossible to know the exact Voat user base since Voat never shared the complete list of user profiles, even when it supported a data API service; to collect a user's profile, one needs to know the username.
This means that we cannot acquire user profile data of ``stalkers.''
Assuming the total known number of usernames is $155K$, we estimate that about $27.1\%$ of the total users' profile data ($41.6K$) is either missing, or deleted profiles.
However,~\cite{papasavva2021qoincidence} show that $13\%$ of the $15K$ users being active in QAnon discussions deleted their profiles. 
Considering that many usernames were deleted every day on Voat, we estimate that this dataset offers the best representation of Voat's user base to date.
Incorporating~\cite{papasavva2021qoincidence} user data with ours, we find $5K$ additional user and $1$ subverse profiles.
The final dataset presented and released with this work is detailed in Table~\ref{tbl:final_dataset}.

\descr{Fair Principles.} 
The data released and presented in this paper aligns with the FAIR guiding principles for scientific data, as described below:\footnote{\url{https://www.go-fair.org/fair-principles/}}
\begin{compactitem}
\item{\emph{Findable}: We assign a unique constant digital object identifier (DOI) to our dataset\cite{zenodo}.\footnote{\url{10.5281/zenodo.5841668}}}
\item{\emph{Accessible}: Our dataset is openly accessible.}
\item{\emph{Interoperable}: We use JSON format to store our dataset since it is widely used for storing data and can be used in various programming languages. 
We also provide a detailed description of our dataset's format in Section~\ref{sec:data_description}.}
\item{\emph{Reusable}: We provide all the available metadata along with our dataset and we extensively document them in this paper, in Section~\ref{sec:data_description}.}
\end{compactitem}

\descr{Ethical Considerations.}
The data collected, presented, and released with this paper are available on the Wayback Machine and also used to be accessible (without the need of a registered account) on Voat before it went down.
The collection and release of this dataset does not violate Voat's or Wayback Machine's Terms of Service.
Although some subverses on Voat allowed users to post anonymously, the overwhelming majority did not offer this functionality.
Hence, we collect user profile data of $114K$ users.
The only identification of these user profiles is the unique pseudo name, which is not personally identifiable information.
Analysis of the activity generated on Voat to other services could potentially be used to de-anonymize users.
We note that we followed standard ethical guidelines~\cite{rivers2014ethical} and made no attempt to de-anonymize users.

\section{Data Description}\label{sec:data_description}
We now present the structure of our dataset, available at~\cite{zenodo}.

Our dataset consists of submission, comment, user profile, and subverse profile data.
We release our data in various newline-delimited JSON files ({\tt.ndjson}).\footnote{\url{http://ndjson.org/}}
Each line in a {\tt.ndjson} file consists of a JSON object that holds various keys and values.
Specifically, we release $7,616$ {\tt.ndjson} files, one for every subverse, that hold the submission data.
Similarly, we release $7,515$ {\tt.ndjson} files that have comment data.
We inspect our dataset for the missing $101$ subverses' comments and find that these subverses have no comment activity, only a small number of submissions.
Also, a single {\tt.ndjson} file is released for user profile data, and another for subverse profile data.
In total, we release $15,133$ {\tt.ndjson} files.
Table~\ref{tbl:data_structure} lists the keys, value data type, and description of our dataset files.

We choose to release the submission and comment data separately for every subverse as we believe it facilitates researchers that want to focus on specific communities.
We also use JSON to release our dataset as it is among the most optimal ways to store and share data as it has extensive documentation and is supported by all popular programming languages.\footnote{\url{https://www.loc.gov/preservation/digital/formats/fdd/fdd000381.shtml}}

\begin{table}[t]
\centering
\resizebox{1\columnwidth}{!}{
\begin{tabular}{l r r}
\toprule
\textbf{Key} & \textbf{Value data type} & \textbf{Description}\\
\midrule
\multicolumn{3}{c}{\tt{subverse\_name\_submissions.ndjson} (7,616 files)}\\
\midrule
\emph{title} & \tt{string} & Title of the submission \\
\emph{body} & \tt{string} & The text posted along with the submission \\
\emph{user} & \tt{string} & Username of the submission creator \\
\emph{time} & \tt{string} & Time the submission was posted \\
\emph{date} & \tt{string} & Date the submission was posted \\
\emph{upvotes} & \tt{integer} & Number of upvotes \\
\emph{downvotes} & \tt{integer} & Number of downvotes \\
\emph{domain} & \tt{string} & Domain the submission links to \\
\emph{link} & \tt{string} & Unique submission URL\\
\emph{submission\_id} & \tt{integer} & Unique submission id \\
\emph{subverse} & \tt{string} & Full name of the subverse \\
\toprule
\multicolumn{3}{c}{\tt{subverse\_name\_comments.ndjson} (7,515 files)}\\
\midrule
\emph{body} & \tt{string} & Comment content \\
\emph{user} & \tt{string} & Username of the comment creator \\
\emph{time} & \tt{string} & Time the comment was posted \\
\emph{date} & \tt{string} & Date the comment was posted \\
\emph{upvotes} & \tt{integer} & Number of upvotes \\
\emph{downvotes} & \tt{integer} & Number of downvotes \\
\emph{comment\_id} & \tt{integer} & Unique comment id \\
\emph{depth} & \tt{integer} & Tree depth level of the comment \\
\emph{subverse} & \tt{string} & Full name of the subverse \\
\emph{root\_submission} & \tt{integer} & Submission id the comment belongs to \\
\toprule
\multicolumn{3}{c}{\tt{user\_profiles.ndjson} (1 file)}\\
\midrule
\emph{user} & \tt{string} & Username \\
\emph{reg\_date} & \tt{string} & Registration date \\
\emph{moderates} & \tt{list of strings} & Subverses the user moderated \\
\emph{owns} & \tt{list of strings} & Subverses the user created \\
\toprule
\multicolumn{3}{c}{\tt{subverse\_profiles.ndjson} (1 file)}\\
\midrule
\emph{subverse} & \tt{string} & The full name of the subverse \\
\emph{subscriber\_count} & \tt{integer} & Number of subscribers \\
\emph{about} & \tt{string} & Description of the subverse \\
\emph{date\_created} & \tt{string} & Date the subverse was created \\
\toprule
\end{tabular}
}
\caption{Description of the dataset files keys and data value types.}\label{tbl:data_structure}
\end{table}

\section{Data Analysis}\label{sec:analysis}
In this Section we provide some statistical analysis and visualization of our dataset.

\begin{figure*}[t]
    \centering
    \includegraphics[width=0.9\textwidth]{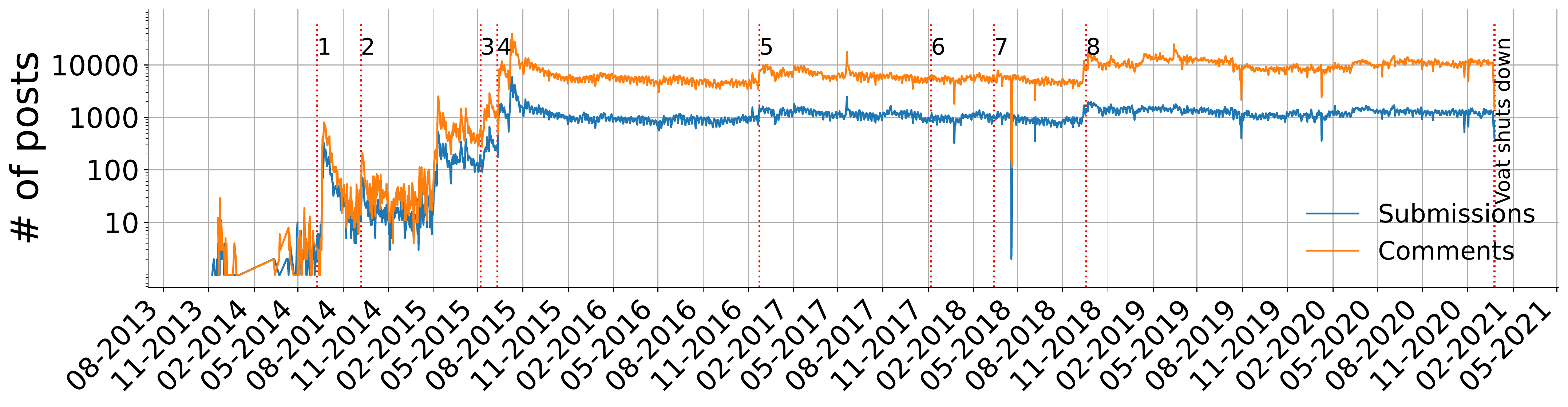}
    \caption{Number of all submissions and comments per day on Voat. Note log scale on y-axis.}
    \label{fig:submission-comment-perday}
\end{figure*}

\begin{figure*}[t]
\centering
\subfigure[Submissions]{\includegraphics[width=0.9\textwidth]{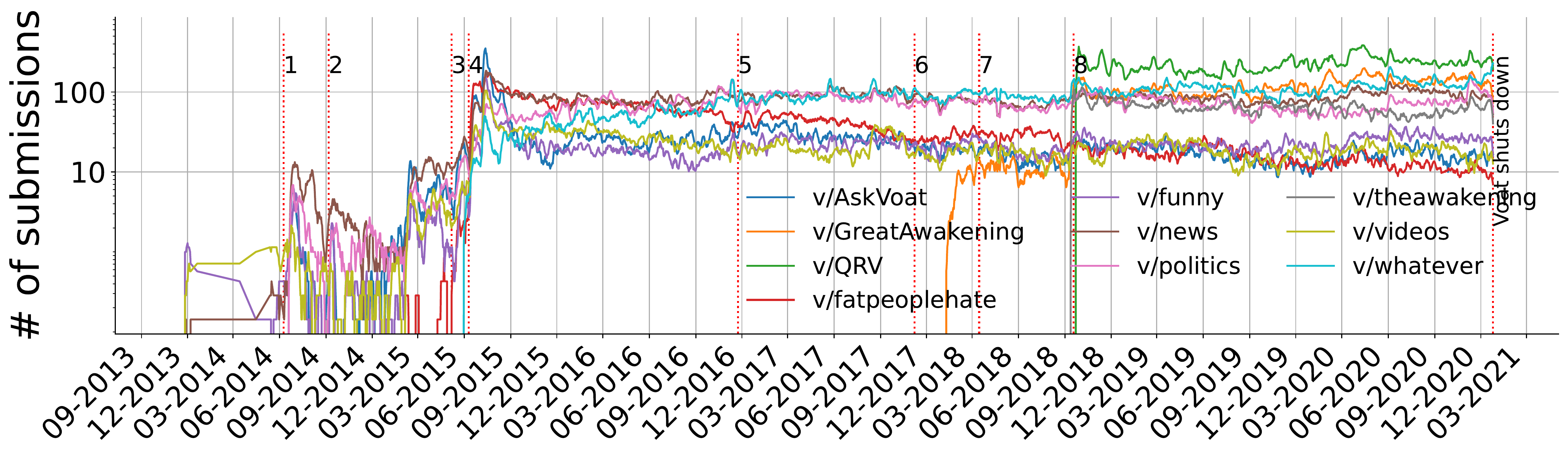}\label{fig:fig:submissions_per_day_top10}}
\subfigure[Comments]{\includegraphics[width=0.9\textwidth]{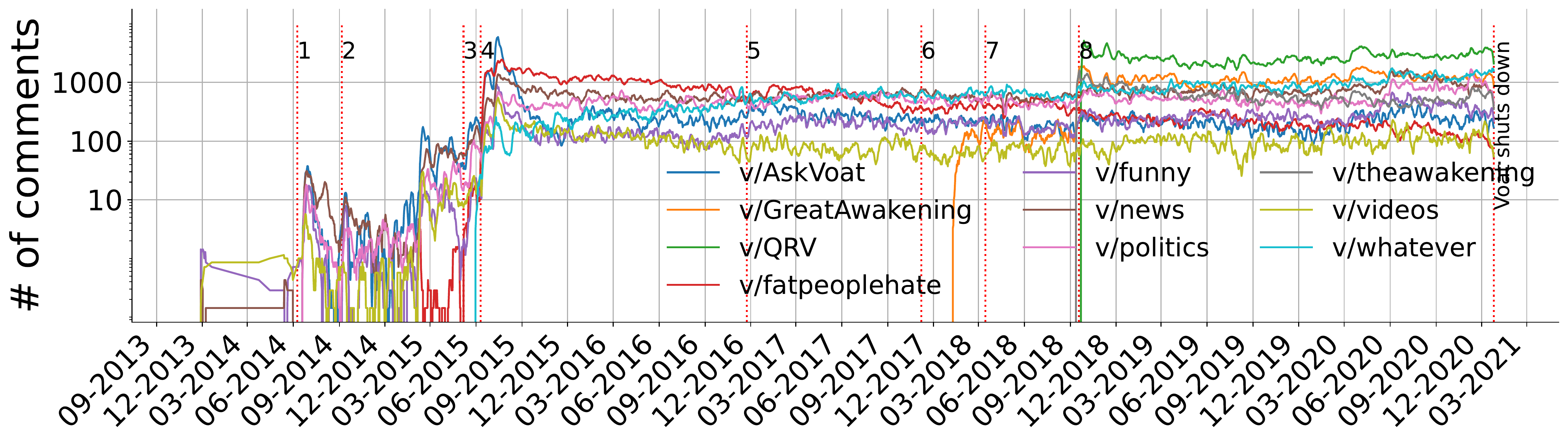}\label{fig:comments__per_day_top10}}
\caption{Seven day average number of a) submissions and b) comments per day on the top 10 most subscribed subverses on Voat. Note log scale on y-axis.}
\label{fig:submission-comment-perday-top10}
\end{figure*}

\descr{Posting Activity.}
First, we show the overall posting activity on Voat.
Figure~\ref{fig:submission-comment-perday} shows the number of submissions and comments per day on the platform.
The vertical red dotted lines represent the events listed in Table~\ref{tbl:events}.
Although the platform was officially launched in April 2014, the first-ever submission was posted by \emph{@Atko} on November 8, 2013, on the \dev subverse.
This subverse was discussing the development of Voat, and at the time, only seven users were posting on the platform.

The total number of submissions in 2013 is only $61$.
These submissions primarily include discussions of @Atko and @PuttItOut in the \dev subverse.
When the platform was launched in 2014, the total number of submissions peaks to $5,268$, then $276K$ in 2015, $324.8K$ in 2016, $397.2K$ in 2017, $382.9K$ in 2018, $439.2K$ in 2019, and for the last year, 2020, $421K$ submissions. 
Overall, there was no significant increase in activity on the platform after 2016.
The most active day on the site is July 10, 2015, with $5.5K$ submissions.
Manual inspection of our dataset indicates that discussions on that day focuses on Donald Trump, vaccine legislation, Reddit's CEO Ellen Pao resigning, and other world happenings.

The date with the most submissions on the site is very close to the day Reddit banned many communities like \rfph and \rnigger~\cite{verge2015migration}.
Shortly after Reddit banned these communities, Voat experienced heavy traffic and downtime~\cite{voatdown2015}.
Regarding comment activity, only $99$ comments were posted in 2013, $13K$ in 2014, $1.6M$ in 2015, $1.8M$ in 2016, $2.1M$ in 2017, $2.4M$ in 2018, $3.8M$ in 2019, and $3.3M$ in 2020
Again, the date with the most comments on the platform is July 10, 2015, with $37.5K$ comments.

In addition, we show the overall activity on Voat in the top ten most subscribed subverses, namely, \askvoat, \greatawakening, \qrv, \fph, \funny, \news, \politics, \theawakening, \videos, and \whatever, in Figure~\ref{fig:submission-comment-perday-top10}.
We present this analysis to show how active the most popular subverses on Voat were since we believe that researchers interested in our dataset might consider these findings useful.
The vertical red dotted lines on the figure indicate the bans listed in Table~\ref{tbl:events}.
When Reddit refugee crowd joined Voat (ban number 1, 3 and 4 from Table~\ref{tbl:events}) many general discussion subverses like \askvoat, \news, \politics, \videos, \funny, and \whatever became more active, indicating that this new influx of users bolstered the overall activity on the platform.

Interestingly, not all banned subreddits appeared on Voat as subverses shortly after a Reddit ban frenzy.
The subverse \greatawakening was created on January 1, 2018, nine months before Reddit banned QAnon subreddits (ban no. 8 ).
This subverse was the 10th most popular subverse when Voat shut down.
QAnon discussion on the platform boomed when \theawakening and \qrv first appeared on Voat on September 12 and September 22, 2018, respectively, with approximately 200 submissions per day on \qrv alone.
These three subverses turned out to be among the top 5 most active subverses on the platform, with \qrv being the most active in both daily submissions and comments on the whole Voat, within only ten days after being banned from Reddit~\cite{awakening2018ban}. 

The figures discussed in this subsection support the reports that Voat was among the main hubs for Reddit migrating communities.
In addition, Figure~\ref{fig:submission-comment-perday-top10} shows that other than general discussion subverses, the most subscribed subverses focused on hate speech (\fph) and conspiracy theories (\qrv, \theawakening, \thegreatawakening).

\begin{figure}[t]
    \centering
    \includegraphics[width=0.8\columnwidth]{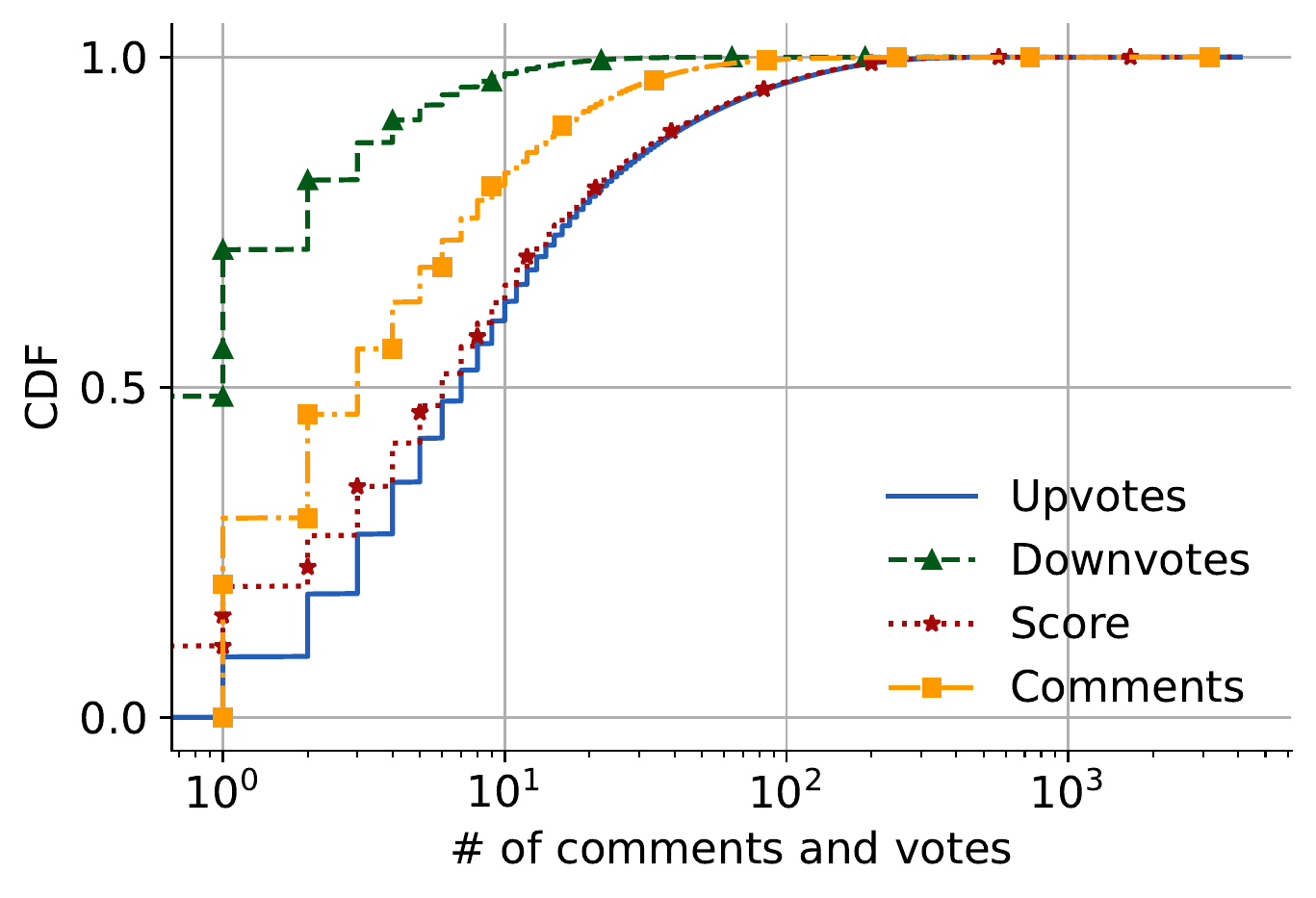}
    \caption{CDF of the number of comments, upvotes, downvotes, and net votes per submission.}
    \label{fig:cdf_comments_votes}
\end{figure}

\descr{Submission Engagement.}
We set to discuss the engagement of the users on the platform.
In Figure~\ref{fig:cdf_comments_votes} we plot the Cumulative Distribution Functions (CDF) of the number of comments, upvotes, downvotes, and net votes (upvotes minus downvotes) per submission.

Submissions on Voat get a median number of $3$ comments, $7$ upvotes, $1$ downvotes, and a net score of $7$.
Comments receive a median $1$ and $0$ upvotes and downvotes respectively.
The most upvoted submission reached over $4K$ upvotes, posted by Atko in \announcements in July 2015, explaining that Voat is experiencing heavy traffic because of Reddit bans and they are working on fixing the issues.
The most downvoted submission ($392$ downvotes) was posted in \politics with the title ``Dear Media: Please Stop Normalizing The Alt-Right.''
The most liked comment noted that ``someone isn't happy that Voat is succeeding'' and reached $1.5K$ upvotes on a submission posted by Atko that was discussing the DDoS attacks Voat was experiencing in July 2015. 
Last, the most disliked comment received $247$ downvotes from a user that was asking @PuttItOut to reconsider the voting system of the site since they lost their submission posting privileges because of people downvoting them when posting their honest opinion.
The user asks the CEOs:
\begin{quote}
\emph{[...]ask yourself: Are you fine with a website that caters to some of the most dangerous people currently walking the planet? Take a look at how depraved Trump supporters are, and ask yourself if free speech is worth the cost:[...]}
\end{quote}

\begin{figure*}[t]
    \centering
    \includegraphics[width=1\textwidth]{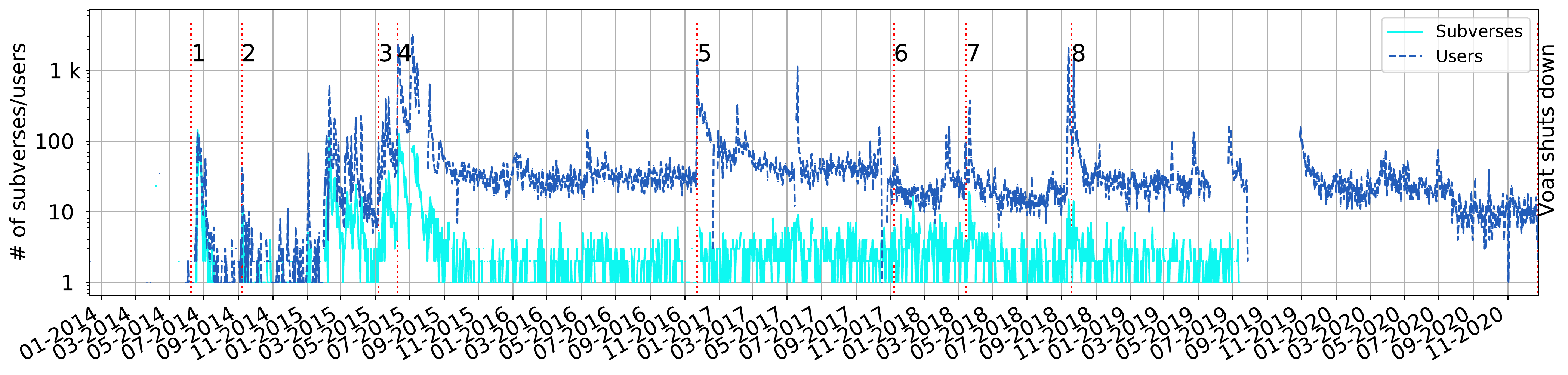}
    \caption{Number of users and subverses registered per day. Note log on y-axis.}
    \label{fig:users_subverses_per_day}
\end{figure*}

\descr{User registration and Subverse creation.}
In Figure~\ref{fig:users_subverses_per_day} we plot the number of daily user and subverse registrations on Voat.
The vertical dotted lines mark the bans listed in Table~\ref{tbl:events}.

The first Reddit ban that seemed to have influenced Voat's user base is the one of  \rbeatingwomen, on June 9, 2014~\cite{beatingwomen} (ban no. 1).
Eleven days after the ban, on June 20, Voat had 145 new subverses in a single day. 
Specifically, the day with the most subverses ever created on the platform.
On June 22, there were 112 new subverses.

Moving on to 2015, we find that July 7 is the date with the most users ever registered on Voat in a single day, $3,175$ registrations, followed by July 5 with $2,854$ registrations.
These dates are close to the date Reddit banned various hate subreddits like \rnigger and \rfph (ban no. 3 and 4).
Also, during summer of 2015, Reddit changed their free speech and content policy~\cite{redditpolicy} and the founder noted that ``Reddit was not created to be a bastion of free speech.''
On July 12 and 13, the platform marked two of the five days with the highest new subverses created, 125 and 112, respectively.
The fifth top date with the most user registrations on Voat is September 13, 2018, with $2,021$ users, probably due to Reddit banning QAnon focused subreddits (ban no. 8).

This analysis provides a glance at Voat's user base and subverse changes over the years.
It is apparent that Reddit influenced Voat activity and that the platform was among the preferred Reddit alternatives for banned users.

\descr{Links.}
Since Voat is a news aggregator platform, we also analyze the domains the users posted on the site to show what kind of content the userbase of Voat consumed.

\begin{figure*}[t]
    \centering
    \includegraphics[width=0.7\textwidth]{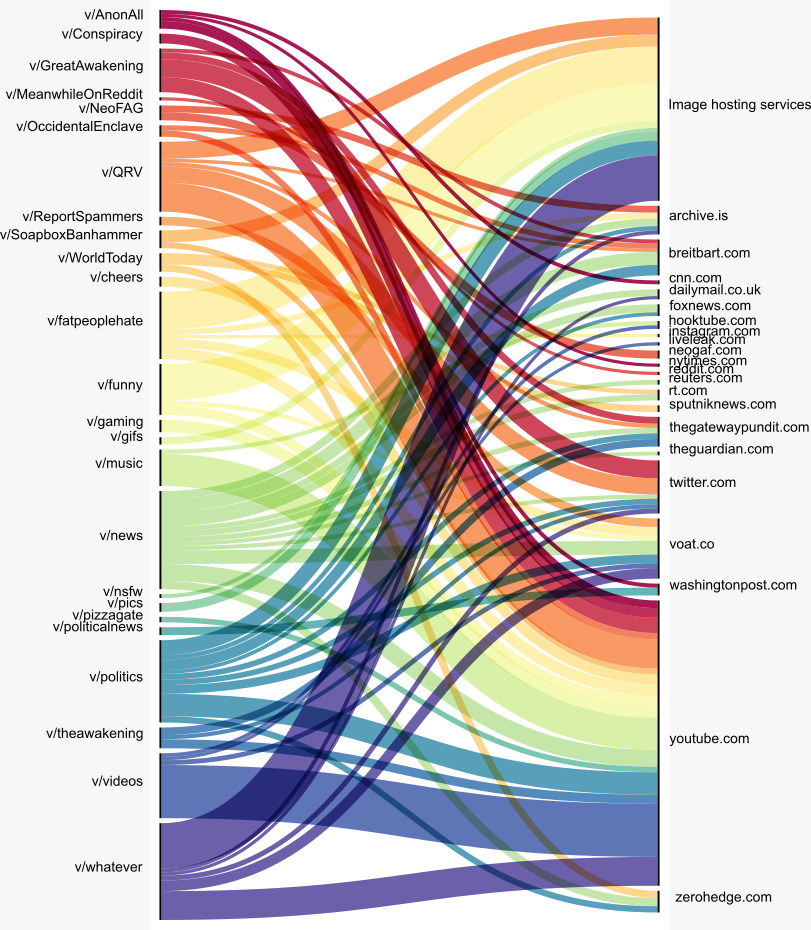}
    \caption{Voat's shared domains ecosystem.}
    \label{fig:links}
\end{figure*}

For each submission that redirects users to other domains, we retrieve the name of the subverse the submission is posted in and the external link it redirects to.
We count how many times a domain is shared in a community, keeping only the subverse and domain pairs that are the most recurrent in the dataset.
The results of this analysis are displayed in Figure~\ref{fig:links}, an alluvial diagram, where the line thickness represents the number of times the domain was shared on the subverse it points to.

Most of the links that redirect users to Reddit were posted in \meanwhileonreddit.
The subverse focusing on body-shaming, \fph, redirected users to Instagram, YouTube, and image sharing services; websites where users can upload images and share the link to that image on other platforms.
The \news subverse linked YouTube, Voat, online press outlets, and archiving services links.
It is known that users in alt-right social networks avoid sharing the direct link to a website and prefer an archive link instead to avoid monetizing the website~\cite{zannettou2018understanding}.
The majority of the alternative news links (Breitbart, GatewayPundit, and Zero Hedge) are posted on \news and \world.
Most of the Twitter links on the website were posted in \qrv and \greatawakening.
Most of the tweets include Donald Trump's tweets and other political discussions on Twitter.

Overall, Voat users shared links to other social networks like 4chan, Twitter, and Instagram.
News on the website was shared via legitimate online press outlets and other alternative news outlets, along with archiving services links.
Most of the images on the platform were shared on \funny, \fph, and \whatever.

\descr{Content Creators.}
We now take a deep look into Voat's user ecosystem.
We attempt to show how users form clusters based on the subverses they most often engaged with (posted a submission or a comment) to show whether the userbase of Voat is homogeneous or not.
Further analysis on Voat's user base may shed light on what content users prefer to see on Voat and whether all of Voat's subverses focused on hateful and politically incorrect content.

As shown from~\cite{papasavva2021qoincidence,amin2021hatemail}, some users are responsible for a large amount of content being shared in some communities, leading to imbalances, influencing the content users consume on the platform.
By analyzing each user's interactions on Voat, we hope to obverse how all these various communities blended after a mass migration from Reddit, or if Voat was nothing more than an aggregate of small, selective echo chambers.
This way, researchers interested in our paper can know what to expect.

\begin{figure*}[t]
    \centering
    \includegraphics[width=0.8\textwidth]{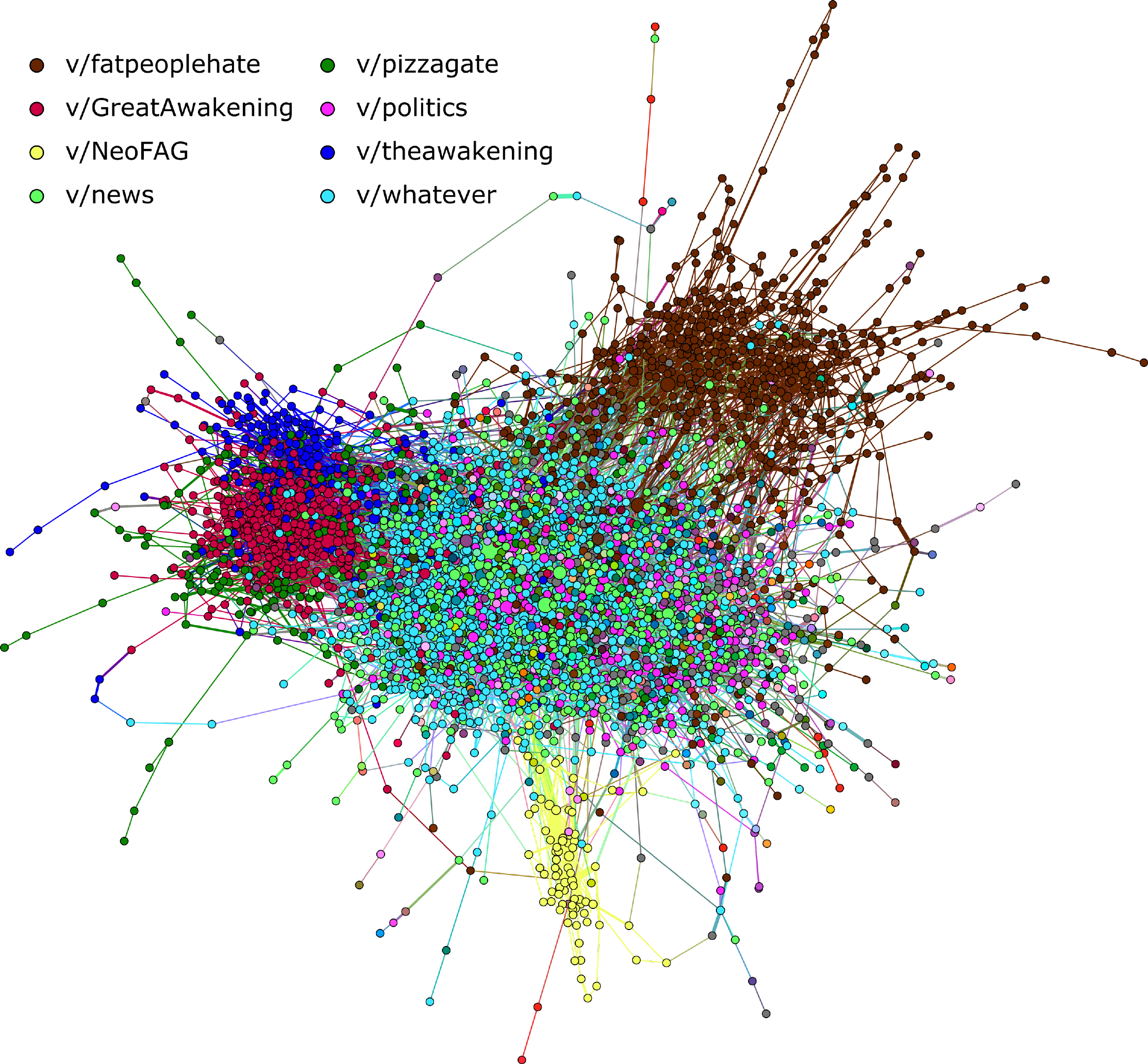}
    \caption{User and subverse interaction ecosystem on Voat.}
    \label{fig:user_interactions}
\end{figure*}

In Figure~\ref{fig:user_interactions} we plot a graph network where nodes represent users, and the edges symbolize their interactions.
For example, users are linked together if they participated in the same conversation, i.e., they both commented on the same submission, or one of them is the submitter while the other one commented.
The weight of the edge is given by the number of interactions shared by the same two users,  and the color represents the subverse where the user participated the most.

The network is composed of a giant cluster, where most of the subverses are mixed together.
This cluster includes \politics, \news, and \whatever, which makes sense since these are general discussion subverses, and it is expected that many users meet there for general discussion.
However,  some of the subverses are strongly isolated in the network.
For example, the \neofag (yellow) community shows that most users tend to only engage within that subverse.
Similarly, \greatawakening (red) and \theawakening (dark blue) seem to be clustered together and somewhat interacting with \pizzagate (dark green).
Some users that engage with these three subverses also engage in the general discussion subverses, which is aligned with the findings of~\cite{papasavva2021qoincidence}.
Last, \fph (brown) users also seem to form their own cluster while infiltrating the general discussion subverses.

To measure the homophily of these communities, we used the $EI$ homophily index, which is a metric that indicates how many members of a network favor in-group interactions rather than out-group ones.
Given a specific node with \textit{E} external edges, i.e., edges with nodes from the out-group, and \textit{I} internal edges, i.e., edges with nodes from the in-group, the $EI$ homophily index is given by the equation $EI = (E-I)/(E+I)$.

An index $EI = +1$ indicates that the node only interacts with members of the out-group,  whereas $EI = -1$ applies to nodes that only interact within their in-group. 
Table~\ref{tbl:homophily} lists the average $EI$ homophily index of the members of the subverses highlighted in the legend of Figure~\ref{fig:user_interactions}.

\begin{table}[H]
\small
\centering
\begin{tabular}{lc}
\toprule
\textbf{Subverse} & \multicolumn{1}{l}{\textbf{EI-Homophily Index}} \\ 
\toprule
politics          & 0.50                                            \\
news              & 0.40                                            \\
whatever          & 0.23                                            \\
theawakening      & 0.01                                            \\
GreatAwakening    & -0.25                                           \\
pizzagate         & -0.49                                           \\
fatpeoplehate     & -0.61                                           \\
NeoFAG            & -0.74                                         \\
\toprule 
\end{tabular}
\caption{Average homophily index between subverses and members.}
\label{tbl:homophily}
\end{table}

Users who are very active on popular subverses such as \politics and \news have a high average $EI$ homophily index, meaning they mostly interact with users from the out-group.
The opposite can be said for subverses like \theawakening, \greatawakening, \pizzagate, and especially, \neofag and \fph.
These communities do not converse a lot outside of their social group.
The $EI$ index is almost zero for \theawakening, meaning users from this community interact as much with the out-group as with the in-group. 
By looking at the community, this can be explained by the fact that users from the communities gravitating around the QAnon narrative, i.e., \theawakening, and \greatawakening, are more connected than other communities.
As a result, the external edges can be nothing more than crossovers between these two subverses.
The userbases of \neofag and \fph seem to be the ones that only prefer to interact with members of their community.

We present this analysis to motivate researchers studying user interactions and echo chambers.
Further research using our dataset may shed light on whether Voat was a bastion of echo chambers or not, along with what narratives users within these communities exchanged.

\section{Related Work}\label{sec:related_work}
In this section, we present existing work focusing on Voat, and other dataset papers similar to ours.
Voat attracted the interest of researchers over the past years, especially after Reddit started banning communities in 2015.
Although some papers mention that their dataset is available upon request, these datasets only include data from a couple of subverses that cover a short period of time.
To the best of our knowledge, our Voat dataset is 1) the only one to be openly and publicly available online, and 2) the most complete and largest one, covering the whole history of Voat, along with data of the users that ever posted a submission or a comment on the platform. 

\descr{Voat research.}
Newell et al.~\cite{newell2016user} collect data from various platforms, including Voat and Reddit and perform, among others, computational analysis to identify the primary motivations that drive users to move to other platforms. 
Chandrasekharan et al.~\cite{chandrasekharan2017bag} collect data from 4chan, Reddit, MetaFilter, and Voat and build a model to detect abusive content online. 
The Voat subverses used in this work include \coon, \nigger, and \fph, all focused on hate towards individuals of specific body or race characteristics, created on Voat shortly after the 2015 Reddit bans~\cite{verge2015migration}.
Similarly, Saleem et al.~\cite{saleem2017web} collect data from Reddit, Voat, and three online forums to train a classifier that detects hateful speech.
Their Voat dataset includes data from \coon, \fph, and \redpill.
A study on deepfakes finds that pornographic deepfakes are mainly created for circulation within the community~\cite{popova2019reading}.
The study uses data from Voat's \deep and the site mrdeepfakes.com, which both were created after Reddit banned the subreddit \rdeep in 2018~\cite{verge2018deep}.

Khalid and Srinivasan~\cite{khalid2020style} compare the features of 872K comments from \politics, \television, and \travel, to Reddit and 4chan comments building a classifier that predicts the origin of the comments based on its style and content.
Papasavva et al.~\cite{papasavva2021qoincidence} collect $0.5M$ posts from \greatawakening, \news, \politics, \funny, and \askvoat to provide an empirical exploratory analysis of the
QAnon community on Voat.
They find, among other things, that \greatawakening is not as toxic as the general discussion subverses. 
Last, Papasavva et al.~\cite{aliapoulios2021gospel} compare Voat's \greatawakening and \news posts to 4chan, 8kun, Reddit, and Q drops (posts posted by ``Q,'' the mastermind behind the QAnon conspiracy theory) on a large scale study on QAnon.
They find that Voat posts are as threatening as Q drops and that content creators on Reddit and Voat only consist of a small portion of the total community. 

\descr{Other datasets.}
One of the largest Reddit datasets is the one of Baumgartner et al.~\cite{baumgartner2020pushshift}, which presents an archiving platform that collects Reddit data and makes them available to researchers since 2015.
The same platform also published over $27.8K$ channels and $317M$ messages from $2.2M$ users from Telegram~\cite{baumgartner2020telegram}.
Fair and Wesslen~\cite{fair2019shouting} release a dataset of $37M$ posts, $24.5M$ comments, and $819K$ user profiles collected from Gab.
Aliapoulios et al.~\cite{aliapoulios2021early} published a dataset consisting of $183M$ posts and $13.25M$ user profiles from Parler, a Twitter alternative.
Last, Papasavva et al.~\cite{papasavva2020raiders} present a dataset with over $3.3M$ threads and $134.5M$ posts from the Politically Incorrect board (/pol/) of the imageboard forum 4chan.

\section{Conclusion}\label{sec:conclusion}
In this work, we present and release a Voat dataset comprising more than $2.38M$ submissions and $16.2M$ comments posted from $113K$ users in over $7K$ Voat subverses.
We combine data collected from Voat API and IAWM released archives to complete the dataset to the best of our ability.
Voat shut down on December 25, 2020, and its data are now otherwise inaccessible.
In this work we also perform a preliminary analysis of the released dataset so researchers interested in it can know what to expect.

Overall, we hope this work further motivates and assists researchers focusing on deplatforming and how users organize massive immigration to other platfroms.
In addition, our dataset could also help answer numerous questions about how `free-speech' sites operated, e.g., do moderators ban users that express opinions other than the ones aligned with the narratives of a subverse? How do other users vote and how toxic are they towards such content? Do sites like these incentivize users to form echo chambers? What kind of content users in this communities consume, etc.?
Also, our dataset could assist multi-platform studies to understand similarities and differences of different communities.
Last, since Voat was a bastion of free-speech, we are confident that access to our dataset could assist researchers towards training algorithms in natural language processing and detecting hate speech, fake news dissemination, conspiracy theories, etc. 
Finally, other than quantitative work, we hope that the data can also be used in qualitative work studying specific events, social theories, and communities.

\section*{Acknowledgments}
This work was partially funded by the UK EPSRC grant EP/S022503/1 that supports the UCL Centre for Doctoral Training in Cybersecurity.
Any opinions, findings, and conclusions or recommendations expressed in this work are those of the authors and do not necessarily reflect the views of the UK EPSRC.

\bibliographystyle{abbrv}
\bibliography{references}
\end{document}